\documentclass[12pt]{article}
\usepackage{psfig}
\newcommand{\g}[1]{\gamma^{#1}}  
\newcommand{\pp}{\lefteqn{\partial}{\mbox{\large /}}{}}
\newcommand{\A}{\lefteqn{\cal A}{\,\mbox{\large /}}{}}
\newcommand{\D}{\lefteqn{D}{\,\mbox{\large /}}{}}
\newcommand{\re}[1]{(\ref{#1}){}}   
\newcommand{\zm}{{\mathrm{z.m.}}}        
\newcommand{\nzm}{{\mathrm{non\ z.m.}}}  
\newcommand{\Det}{\det(i\pp + {\A})}         
\newcommand{\F}{{\cal F}}          
\newcommand{\anom}{{\mbox{\textit{\Large a}}}}
\newcommand{\df}{\delta^2(x_\perp)}
\newcommand{\pr}{\prime}          
\newcommand{\xp}{x^+}
\newcommand{\xpp}{x^{\pr +}}
\newcommand{\pxp}{\partial_+}
\newcommand{\pxm}{\partial_-}
\newcommand{\jp}{j_+}
\newcommand{\uab}{U_{\alpha\beta}}
\newcommand{\dfrac}[2]{\displaystyle\frac{#1}{#2}}

\newcommand{\Tr}{\mathop\mathrm{Tr}}    

\begin{document}

\hfill\vbox{\hbox{hep-th/0007037}\hbox{EFI-2000-18} }\break \vskip 2.0cm

\centerline{\large \bf The Local Structure of Anomaly Inflow}

\vspace*{6.0ex}

\centerline{\large \rm Jeffrey A. Harvey and Oleg Ruchayskiy}

\vspace*{6.5ex}

\centerline{\large \it Enrico Fermi Institute and Department of Physics}
\centerline{\large \it University of Chicago, Chicago, IL 60637}
\vspace*{1ex}
\centerline{\it harvey@theory.uchicago.edu, ruchay@flash.uchicago.edu}
\vspace*{4.5ex}

\begin{abstract}
  
  Anomaly cancellation for M-theory fivebranes requires the introduction of a
  ``bump-form'' which smoothes out the five-brane source. We discuss the
  physical origin of this bump-form in the simpler case of axion strings in
  $3+1$ dimensions and construct it in terms of the radial profile of the
  fermion zero modes. Our treatment allows for a clearer understanding of the
  role played by covariant rather than consistent anomalies when anomalies are
  canceled by inflow from the bulk.  We briefly discuss the generalization of
  these results to fivebrane anomalies in M theory.
\end{abstract}

\vfill \eject

\section[Introduction]{Introduction} 

Anomalies are known to be an important tool for getting non-perturbative
information about quantum field theories. The requirement of cancellation of
anomalies in local gauge symmetries gives important constraints on the
structure of the theory, and anomalies in global symmetries often give
important non-perturbative information about the theory.

In the case of M-theory where we lack a formulation that is both complete and
practical for computation, the study of anomalies is especially important as a
means of obtaining exact results.  An analysis of diffeomorphism anomalies in
the presence of a M5-brane (\cite{Duff,Witten,FHMM,Bonora:1999ie,Becker}),
shows that the ``standard'' mechanism of cancellation of anomalies by inflow
from the bulk does not work in the expected way.  Namely, there remains a
\emph{normal bundle} anomaly, coming from the part of the diffeomorphism group
which acts on the fermion zero modes as $SO(5)$ gauge transformations by
rotating the normal bundle at each point on the brane. There were a number of
attempts to treat this problem. In \cite{Witten} it was shown that if the
theory is compactified on a circle (i.e. going to Type IIA string theory) then
the anomaly can be canceled by a local counterterm on the brane. In
\cite{Bonora:1999ie} a local counterterm was suggested for the
eleven-dimensional case (with the result of~\cite{Witten} restored upon
compactification).  However this work is based on the fact that the
world-volume of 5-brane is closed, being the boundary of some
seven-dimensional manifold. This cannot be considered to be a totally
satisfactory answer, as branes can be taken to have boundaries and as there is
no physical meaning so far to this seven dimensional manifold.

 The work \cite{FHMM} was based on the idea that anomaly cancellation in
the presence of the Chern-Simons term in eleven-dimensional supergravity
requires smoothing out the M5-brane source. By smoothing the brane,
modifying the Chern-Simons term in the bulk action, 
it was shown that the anomaly of the normal bundle
indeed cancels. The relation of this mechanism to the cancellation in type
IIA theory was discussed in \cite{Becker}.  The price paid for this anomaly
canceling mechanism   was  the presence in the action of
the \emph{bump function}: an arbitrary function of the distance from the brane
with  definite boundary values. So the question remained: what is the physical
origin of the bump 
function $\rho(r)$ which enters the action and is required for anomaly
cancellation?

In what follows we will consider the role of such a bump function in a much
simpler model where its physical origin can be analyzed. It is connected to
the relation between consistent and covariant anomalies and can be related to
the radial profile of fermion zero modes on the brane. Following this analysis
we make some remarks on the extension of these results to the M5-brane. It
should be noted that cancellation of anomalies can be phrased in global
topological terms \cite{agg} for which the present considerations are
irrelevant. Rather, we are interested here in understanding the detailed local
structure of anomaly cancellation from a physical point of view.  Such
considerations might be relevant in applications of anomaly inflow to
condensed matter systems \cite{bdf,sg,volovik} and the description of chiral
fermions in lattice gauge theory \cite{kaplan}.

\section{Axion Electrodynamics}
\label{sec:Axion-Electr}

The example we are going to deal with is  \emph{Axion
Electrodynamics}. As we will see, when analyzed in detail, it possesses
some  features similar
to those of the five-brane case.  The bulk theory is anomalous in the presence
of a  topological defect (a $1+1$ dimensional string in this case), as is the theory
on the defect.  Inflow from the bulk
cancels this anomaly, but a complete understanding of the cancellation
requires a
detailed  treatment of the string source, which takes into account its
profile. We begin with a quick review of material from
\cite{Callan:1985,Naculich} and then go on to discuss the role of the
bump-form and its physical origin.

Consider a  theory in 3+1 dimensions, which describes  Dirac fermions
interacting with a  U(1) gauge field and a  complex scalar field (which is
neutral under the $U(1)$ gauge symmetry). The Lagrangian
is
\begin{equation}
 \label{eq:qed}
  {\cal L} = -\frac 14 F_{\mu\nu}^2 + \bar \psi i\D \psi + 
  |\partial_\mu\Phi|^2 +  g\bar\psi(\Phi_1 + i\gamma^5\Phi_2)\psi - V(\Phi)
\end{equation}
The complex scalar field $\Phi=\Phi_1 + i\Phi_2$ has a potential $ V(\Phi) $ which
we take to have the form 
\begin{equation}
  \label{eq:5}
  V(\Phi) = \lambda (|\Phi|^2 - v^2)^2
\end{equation}
The equations of motion  admit a global vortex  solution of the form:
 \begin{equation}
   \label{eq:profile}
   \Phi(x^\mu) = f(x_\perp)e^{i\theta(x^\mu)}
 \end{equation}
 Here $x_\perp = \sqrt{\strut x^2+y^2}$ is the radial coordinate in the
 $(x,y)$-plane, and the phase $\theta(x^\mu)$ gives the classical value of the
 \emph{axion} field. The profile function $f(x_\perp)$ is assumed to be zero
 at the origin and equal to the $v$ far from it.  For a vortex with winding
 number one the axion field is $\theta(x^\mu) = \varphi$, where $\varphi$ is a
 polar angle in the $(x,y)$ plane.  This configuration (known as an
 \emph{axion string}) describes a topological defect with codimension~2 and
 topological charge one\footnotemark{}.
 \footnotetext{An axion string with  topological charge $n$ would have
   $\theta(x^\mu) = n\varphi$}

\subsection{Fermion Zero Modes and Anomaly Inflow}
\label{sec:zm}

The index theorem guarantees that the Dirac equation in the
background~(\ref{eq:profile}) has a \emph{zero-mode} solution, i.e.
\begin{equation}
  \label{eq:dirac1}
  \left (i\pp + f(x_\perp)e^{i\g5 \varphi}\right ) \psi_\zm = 0
\end{equation}
(we've absorbed the  Yukawa coupling constant $g$ into the function $f$
 throughout the rest of
the paper). The zero mode solution is~\cite{Callan:1985}:
\begin{equation}
  \label{eq:7}
  \psi_\zm = u_\alpha e^{-i p\left( t+ z \right)} \F(x_\perp)
\end{equation}
where we have introduced the function $\F$ which is the radial  profile of
the fermion  zero mode:
\begin{equation}
  \label{eq:8}
  \F(r) = {\cal C} e^{-\int\limits_0^r f(\sigma)
      d\sigma}
\end{equation}
with ${\cal C}$ a normalization constant. The spinor $u_\alpha$ has the form
$u = (1 - i\g1)\eta$. Here spinor $\eta$ obeys $\gamma^{int} \eta = - \eta$
and $\g5\eta = - \eta$ with $\gamma^{int} = \g0 \g3$ and $\g5 = i
\g0\g1\g2\g3$. In the conventions of \cite{Peskin} it has the explicit form
\begin{equation}
  \label{eq:9}
  u_\alpha =  \frac {1} {\sqrt{2}}\left (
    \begin{array}{c}
      1\\
      0\\
      0\\
      -i
    \end{array} \right )
\end{equation}
Note that this solution decays exponentially in the $(x,y)$-plane, being
quasi-two-dimensional and describes a fermion of negative (two-dimensional)
chirality, propagating in the $-z$ direction.  In order to obtain canonically
normalized two-dimensional fermions we impose the normalization condition
\begin{equation}
  \label{eq:11}
  \int d^2 x_\perp\,dz\, \psi^*_{p} \psi_{p^\prime} = 2\pi\delta(p -
  p^\prime) 
\end{equation}
which in particular determines the  normalization constant ${\cal C}$
from the condition: 
\begin{equation}
  \label{eq:12}
  \int d^2 x_{\perp}\, \F^2(x_\perp) = 1
\end{equation}
The theory of fermions of one chirality in two dimensions is anomalous when
coupled to an electromagnetic field\footnotemark{} 
\footnotetext{In this paper  Latin
  indices ($a,b$) run over subset $\{0,3\}$ and Greek indices run $0,\dots,
  3$}
\begin{equation}
  \label{eq:28}
  \partial_a j^a = - \frac e {8\pi} \epsilon^{ab} F_{ab}
\end{equation}
and thus violates charge conservation. In our case, however,
this theory is  naturally embedded into a  non-anomalous 
$3+1$-dimensional theory and charge conservation should be ensured by
anomaly inflow from the bulk. 

Using the ideas of \cite{gw} one can compute the current far from the
string in the presence of a background electromagnetic field by
computing a one loop Feynman diagram with one
insertion of the electromagnetic field and one insertion of the scalar field.
One finds  (\cite{Callan:1985,Naculich}):
\begin{equation}
  \label{eq:current1}
  \langle j^\mu\rangle = \frac {e} {8\pi^2} 
  \epsilon^{\mu\nu\lambda\sigma}\partial_\nu\theta F_{\lambda\sigma} = \frac
  {e} {4\pi^2}\partial_\nu(\theta\tilde F^{\mu\nu}) 
\end{equation}

In the presence of a constant electric background field pointing along the
axis of the axion string this gives a current flowing radially inward and thus
bringing electric charge to the string. This is the anomaly inflow we are
looking for. The divergence of the current on the string can be computed using
the fact that
\begin{equation}
  \label{eq:comm}
 [\partial_x,\,\partial_y]\theta = 2\pi\delta^{(2)}(x_\perp)
\end{equation}
to give
\begin{equation}
  \label{eq:27}
  \partial_\mu \langle j^\mu\rangle = \frac {e} {4\pi} \epsilon^{ab} F_{ab}
  \delta^2(x_\perp)
\end{equation}
twice the expression~(\ref{eq:28})!

This discrepancy was explained in \cite{Naculich} as arising from
the difference between the covariant and consistent anomaly which
for an Abelian gauge theory in $1+1$ dimensions is simply a factor
of two \cite{bz}. The  anomaly~(\ref{eq:28}) is the consistent anomaly in
$1+1$ dimensions, following from variation of a $1+1$ dimensional
action. 
The current~(\ref{eq:current1}) can be obtained from the action:
\begin{equation}
  \label{eq:S_eff}
  S_{eff} = - \frac {e^2}{ 16\pi^2}\int_{M_4}  
  \epsilon^{\mu\nu\lambda\sigma} \partial_\mu\theta A_\nu F_{\lambda\sigma}
\end{equation}
which up to an integration by parts is the usual coupling of the
axion to gauge fields which results from integrating out the fermion
fields. This action is covariant as is the resulting current. 

It was proposed in~\cite{Naculich} that there should be an additional
contribution to the current on the string which converts the covariant current
flowing in from infinity into a covariant current on the axion string. It was
also noted there that such a term could be derived formally by treating the
action (\ref{eq:S_eff}) as valid everywhere, not just far from the string.
Variation of the action then leads to an addition contribution to the current
which is delta-function localized on the string and is of precisely the right
form to convert the consistent current on the string into the covariant
current.

\subsection{Anomaly Cancellation Revisited}
\label{sec:anom-canc-revis}

{}From the analysis in the last section we see 
several reasons why we would like
to better understand the physics of anomaly inflow.  First, the naive analysis
shows that divergence of the current flowing in to the string does not match
the divergence of the current on the string computed in the $1+1$-dimensional
theory. The modification suggested in~\cite{Naculich} extrapolates the
formula~(\ref{eq:S_eff}) to the whole bulk. It is obvious, however, from the
way this formula has been derived that it cannot be valid in the region near
the string.  Second, usage of equations similar to~(\ref{eq:comm}) is hard to
justify rigorously, as neither $\theta$ nor $d\theta$ are well defined at the
origin and its second derivative is ambiguous.
The main reason, however, and the original motivation was the hope that we
will be able to understand better  the procedure used  in~\cite{FHMM} to
smooth out the fivebrane source. In
the current case we are in much better shape though, as both the form of the
original Lagrangian and the fundamental degrees of freedom are known. So, we
may be able to shed some light on the question: ``What is the
$\rho$-function?''

To summarize: we are hoping to derive an
effective action in the bulk with the following properties:
\begin{enumerate}
  
\item It is well defined in the whole $M_4$ -- four-dimensional bulk and not
  only far from the string

\item The divergence of the current following from this action matches
  the divergence of the current from the anomaly of the 
  string zero modes.
  
\item The string source is described as a  smooth solution and the
  delta-function equations~\re{eq:comm} are modified -- that is smoothed out.

\end{enumerate}

Following~\cite{FHMM} we will first postulate an interaction term in the
effective action given by 
\begin{equation}
  \label{eq:S_eff3}
  S_{eff} = -\frac {e^2}{ 16\pi^2}\int_{M_4}
  \epsilon^{\mu\nu\lambda\sigma}(1+\rho) \partial_\mu\theta A_\nu
  F_{\lambda\sigma} 
\end{equation}
Here we've introduced $\rho(r)$, a monotonic function (zero-form) of the
transverse radius $r$ with the properties:
\begin{equation}
  \label{eq:rho1}
  \rho(0) = -1;\quad \rho(r \to\infty) = 0
\end{equation}
This imply in particular that:
\begin{equation}
  \label{eq:rho4}
  \int\limits_0^\infty dr\,\rho^\pr(r) = \rho(\infty) - \rho(0) = 1
\end{equation}
The action~(\ref{eq:S_eff3})  is well-defined everywhere in $M_4$, as
the form $d\varphi$ (in the future we do distinguish  between $d\theta$ and
$d\varphi$, considering the axion field to be in its background value) is
multiplied by $(1+\rho)$, which is zero on the string (at $r=0$).

A function with the properties~\ref{eq:rho1} --- \ref{eq:rho4} can be used to
define a  smooth generalization of the  delta-function describing the 
 embedding of the string world-sheet into spacetime: $\Sigma_2
\hookrightarrow M_4$, similar to  the Poincare dual of the submanifold
(see e.g.~\cite{Bott}), i.e. the form (rather cohomology class)
$\Omega_2^{(P)}$ such that for any 2-form $\omega_2$ on $M_4$:
\begin{equation}
  \label{eq:dual}
  \int_{M_4} \Omega_2^{(P)}\land\omega_2 = \int_{\Sigma_2}\omega_2 
\end{equation}
In terms of the  \emph{bump-form} $d\rho = \rho^\prime(r) dr$ (where $\rho$ has
properties as above), a representative of the  Poincare dual class can be 
written as:
\begin{equation}
  \label{eq:dual5}
  \Omega_2^{(P)} = \frac 1 {2\pi}d\rho\land d\varphi 
\end{equation}
This object does have all the properties we would like for a smoothed
$\delta$-function. From this point on whenever we write $\df$ we will actually
imply its smoothed version~(\ref{eq:dual5}) and will use the $\delta$-function
symbol mostly for convenience.

Varying eq.~(\ref{eq:S_eff3}) one can get the expression for the current:
\begin{equation}
  \label{eq:52}
  j^\mu = \frac{e}{ 8\pi^2}\epsilon^{\mu\nu\lambda\sigma}(1+\rho)\partial_\nu
  \theta F_{\lambda\sigma} + \frac{e}{8\pi^2} \epsilon^{\mu\nu\lambda\sigma}
  (\partial_\lambda \rho)\,( \partial_\nu \theta)\, A_\sigma
\end{equation}
In this case the integration by parts doesn't produce a surface term, as it is
multiplied by $ (1+\rho)$ which is zero at the surface of the string.  The
second term in the eq.~(\ref{eq:52}) can be rewritten as (note that
$\partial_{i}\rho\ne 0$ only for $i=1,2$):
\begin{equation}
  \label{eq:54}
  \Delta j^\mu = -\frac {e}{8\pi^2} \left \{
    \begin{array}{ll}
      0, & \mu \in \{1,2\} \\
      \epsilon^{\mu\nu}\, A_\nu\,  d\rho \land d\varphi, & \mu \in \{0,3\}
    \end{array}
  \right.
\end{equation}
(here we've defined $\epsilon^{\mu\nu} = \epsilon^{ab}\; \mu,\nu= \{0,3\}$, 0
otherwise).  This is almost the same as the additional contribution to the
current~(\ref{eq:current1}) found in~\cite{Naculich}. There it was shown that
to explain discrepancy of the factor of two between anomalies in the bulk
(given by~(\ref{eq:27})) and on the string (eq.~(\ref{eq:28})) an addition to
the current in the form~(\ref{eq:54}) was needed. The full current in our case
will be:
\begin{equation}
  \label{eq:55}
  j^\mu = \frac{e}{ 8\pi^2}\epsilon^{\mu\nu\lambda\sigma}(1+\rho)\partial_\nu
  \theta F_{\lambda\sigma}  - \frac {e}{4\pi}\epsilon^{\mu\nu} A_\nu \df 
\end{equation}
(the only difference with~\cite{Naculich} being the factor $(1+\rho)$ in the
first term and the fact that the delta function in the second
term is now the smoothed out version~(\ref{eq:dual5})). 
The divergence of this current integrated over $M_4$
is 
\begin{equation}
  \label{eq:56}
\int_{M_4}  \partial_\mu j^\mu =
 \int_{M_4} \frac{e}{ 8\pi^2} \epsilon^{\mu\nu\lambda\sigma} 
  \partial_\mu\rho \partial_\nu \theta \partial_\lambda A_\sigma =
\int_{\Sigma_2} \frac{e}{
    4\pi} \epsilon^{ab}\partial_a A_b = \int_{\Sigma_2} \frac{e}{4\pi} F_{03}
\end{equation}
which correctly cancels  the anomaly~(\ref{eq:28}).

\subsection{Computation of the Determinant of the Dirac\\ Operator}
\label{sec:comp-determ-dirac}

We now want to derive~(\ref{eq:S_eff3}) from first principles.
We need in principle to
implement the usual procedure of QFT: start from the partition function
of the full theory (fermions coupled to all the fields which we call
${\cal A}$ for the moment):
$$
Z = \int {\cal D}\psi{\cal D}\bar\psi{\cal D A}\,e^{iS[\psi, {\cal A}]}
$$
and integrate out the fermions to get:
\begin{equation}
  \label{eq:det}
  Z =\int{\cal D A}\,\det(i\pp +{\cal A})=\int{\cal D A}\, e^{i S_{eff}[{\cal
      A}]}
\end{equation}
The determinant of the Dirac operator~(\ref{eq:det}) (if computed exactly)
will then be a well defined gauge invariant object. The effective
action~(\ref{eq:S_eff}), however, was obtained using only expressions for the
wave functions far from the string, i.e. we took into account only part of the
spectrum and so it is not surprising that we find a lack of gauge invariance.
Moreover, the effective action of the form~(\ref{eq:S_eff}) does not contain
any information about the structure of the theory near the string, treating it
as a singular source. Solutions of the Dirac equation, used in the
computations in the previous section which led to~(\ref{eq:S_eff}) were just
plain waves, i.e. the solutions of the eq.~(\ref{eq:dirac1}) in the
topologically trivial background $f(x) = v$.  To obtain the action in a
form~(\ref{eq:S_eff3}) we need to bring some information about the behavior of
the theory near the string.

Said another way, one could try to construct the effective action by doing a
Taylor series expansion of $f(x_\perp)$ and including the radial variations in
a perturbative calculation of the current.  One quickly sees however that this
does not change the current at any finite order in perturbation theory,
essentially because the perturbative current is independent of the magnitude
of $f$.  Instead we must go beyond perturbation theory about a plane wave
basis of fermion states by including the explicit form of the zero mode wave
function in the calculation. That is we would like to 
compute the effective action non-perturbatively in $f(x_\perp)$ while
still in the lowest order in $A_\mu$.

In principle, to obtain the 
effective action in the bulk we would need to compute:
\begin{equation}
  \label{eq:1}
  \Det = \prod \lambda_n
\end{equation}
where $\lambda_n$ are eigenvalues of the Dirac operator in the external field
${\cal A}$:
\begin{equation}
  \label{eq:10}
  (i \pp + \A) \psi_n = \lambda_n \psi_n 
\end{equation}
We are not really interested directly in the determinant. Rather, we
are interested in the current which arises from varying the
determinant with respect to the gauge field and the decomposition
of this current into a part arising from the zero modes, and the rest
which we wish to describe in terms of an effective action. 
We propose to study this decomposition 
in the following way. 

The physical picture we have so far is that the current flowing radially
inward from infinity
toward the string (with  divergence
given by eq.~(\ref{eq:27})) transforms into the current along the string (with
the divergence which should be twice the expression~(\ref{eq:28})). We know,
that the gauge variation of the full $\Det$ is proportional to the divergence
of the current $\partial_\mu j^\mu$. As the full current is conserved, it means
that the  contribution to the divergence of the four-dimensional current coming
from the zero modes only (we will call this part of the current $j_\zm^\mu$)
is opposite to that of coming from the effective action~(\ref{eq:S_eff3}).
Thus by computing the divergence of $j^\mu_\zm$ we will be able to match it
against the gauge variation of (\ref{eq:S_eff3}) and extract the expression
for~$\rho(r)$.

Computation of the zero mode current 
$j_\zm$ is a much easier task by itself.  Zero mode
solutions~(\ref{eq:7}) are the only contributions to it. The reason for this
is very simple -- any other solution of the Dirac equation~(\ref{eq:dirac1})
will be massive from the 1+1 dimensional point of view and will not
contribute to the axial anomaly.

Consider the standard fermion operator 
\begin{equation}
  \label{eq:14}
  \hat \psi(x) = \sum\limits_n \psi_n(x) \hat a_n
\end{equation}
Here $\psi_n$'s are the (exact) eigenfunctions of the Dirac operator in the
background field~(\ref{eq:profile}) and $\hat a_n$ are creation/annihilation
operators of the particles (we will do all this in much more detail in the
next section).  Now, split the r.h.s. of the expression~(\ref{eq:14}) into two
terms: $\hat\psi_\zm$ which is constructed solely in terms of
solutions~(\ref{eq:7}) and $\hat \psi_\nzm$ -- all the other solutions.
The Green's function $G(x, x^\pr) = <\hat\psi \hat\psi^+>$ 
naturally splits then
into $G_\zm(x, x^\pr) + G_\nzm(x, x^\pr)$, where:
\begin{equation}
  \label{eq:6}
  G_\zm(x, x^\pr) = <\hat\psi_\zm(x) \hat\psi_\zm^+(x^\pr)>
\end{equation}
The zero mode contribution to the full four-dimensional current will be: 
\begin{equation}
  \label{eq:21}
  <j^\mu (x) >_\zm = (-ie) (-1) \int d^4 y \Tr \Bigl (\g{\mu} G_\zm (x,y)
  \g{\nu} G_\zm (y,x) \Bigr ) A_\nu (y)
\end{equation}

\subsubsection{Computation of the Current}
\label{sec:current}

\begin{figure}[htbp]
  \begin{center}
    \psfig{file=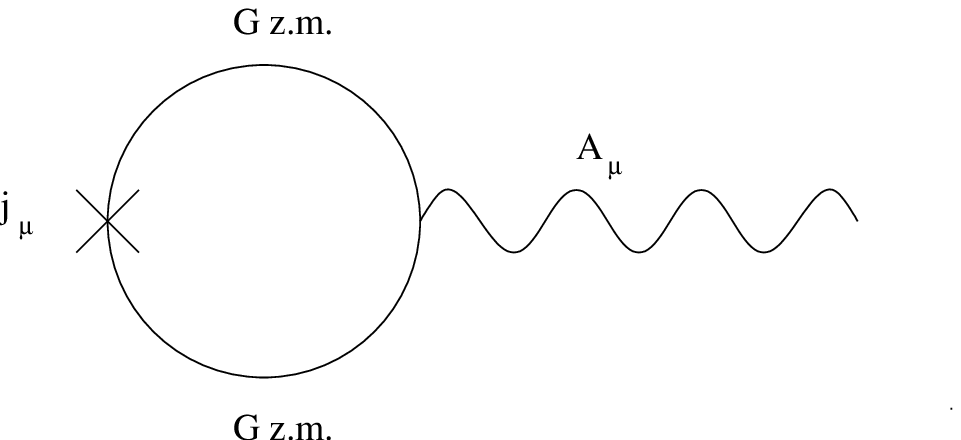}
    \caption{Contribution to the current from zero modes}
    \label{fig:1}
  \end{center}
\end{figure}

Now we will implement in detail the procedure described in the previous
section. First, we construct the Green's function $G_\zm$ of the zero modes in
the background scalar field $\Phi$ (using the function~(\ref{eq:7}) only).
Then using $G_\zm$ we compute the expectation value of the current to lowest
order in the electromagnetic field~(\ref{eq:21}).  This is just the standard
Feynman loop diagram shown on fig.~\ref{fig:1}, where the full Green's
functions $G(x, x^\pr)$ are substituted with zero mode Green's functions
$G_\zm$.  We will read off the gauge variation of the effective action in the
bulk from the divergence of this current.

To construct Green's function we write the $\psi$-operators for the zero mode
solution~(\ref{eq:7}) first. Note, that in the eq.~(\ref{eq:7}) $p > 0$
corresponds to the particle and $p < 0$ -- to the antiparticle, both
propagating in the same direction. Thus we write:
\begin{equation}
  \label{eq:13}
  \hat\psi_\alpha (t,z | r) = \left [\, \int \limits_0^\infty \frac
    {dp}{2\pi}\, u_\alpha\, \hat a_{p}  + \int \limits_{-\infty}^0 \frac
    {dp}{2\pi}\, u_\alpha\, \hat b_{-p}^+ \right ]e^{-i p\left( t+ z \right)}
  \F(r) 
\end{equation}
Here $\hat a_p$ -- annihilates particle with momentum $p$ along the string,
$\hat b_p^+$ -- creates antiparticle with the same momentum.  They obey
standard anticommutation relations. The Green's function is defined by the
usual expression:
\begin{equation}
  \label{eq:17}
  G(x,\,x^\pr) = \left \{
    \begin{array}{rr}
      < \psi(x) \,\bar \psi(x^\pr)>   &\quad t > t^\pr \\
      - < \bar \psi(x^\pr)\, \psi(x)> &\quad t < t^\pr
    \end{array}
    \right .
\end{equation}
From~(\ref{eq:13}):
\begin{equation}
  \label{eq:15}
  <\psi_\alpha (\xp|r)\, \bar \psi_\beta (\xpp| r^\pr)> = \int
  \limits_0^\infty \frac {dp}{2\pi}  \frac {dp^\pr}{2\pi}u_\alpha \bar u_\beta
  e^{-i p\xp} e^{ip^\pr\xpp } < a_p a^+_{p^\pr} > \F(r)\F(r^\pr) 
\end{equation}
(we've introduced standard``light-cone'' coordinates: $x^\pm = t \pm z$) and
analogously for $<\bar \psi_\beta (x^\pr)\, \psi_\alpha (x)> $.  As a results:
\begin{equation}
  \label{eq:20}
  G_\zm (\xp,\, \xpp|r,r^\pr) = \frac 1 {2\pi i}\, ||\uab||\,\F(r)\F(r^\pr)
  \times \frac 1 {\xp - \xpp}
\end{equation}
We see that Green's function factorizes into $(t,z)$ and $(x,y)$ parts, and
that in its $(t,z)$ part it is a standard two-dimensional Green's function of
one chiral fermion. In the above we have defined 
$U_{\alpha\beta} = u_\alpha\otimes\bar u_\beta$.

Now we are in a  position to  compute the expectation value of the
electromagnetic current~(\ref{eq:21}). 
We do the gamma-matrix algebra first:
\begin{equation}
  \label{eq:22}
  \Tr\bigl ( \g{\mu} U \g{\nu} U \bigr ) = \left \{
    \begin{array}{rlcl}
      1,  & \mu = \nu = 0      & \mbox{ or } & \mu = \nu = 3\\
      -1, & \mu = 0,\; \nu = 3 & \mbox{ or } & \mu = 3,\; \nu = 0 \\
      0,  & \mbox{otherwise}
    \end{array} \right .
\end{equation}
Then it immediately follows from eqs.~(\ref{eq:21},~\ref{eq:22}) that
\begin{equation}
  \label{eq:25}
  \begin{array}{lll}
    <j^{1,2}(x)>_\zm & = & 0 \\
    <j^0(x)>_\zm  & = &- <j^3(x)>_\zm 
  \end{array}
\end{equation}
This is in accordance with the fact, that for the fermions of negative
chirality in two dimensions only the $\jp$ component of the current is
non-zero.  Eq.~(\ref{eq:25}) implies that anomaly defined by $ \anom =
\partial_\mu j^\mu $ is:
\begin{equation}
  \label{eq:26}
  \anom  = (\partial_0 - \partial_3 ) <j^0(x)>_\zm  = 2\pxm \jp(x)
\end{equation}
Using the explicit form of the Green's function~(\ref{eq:20}) and taking
$A_\mu$ to be a function of $(t,z)$ it is straightforward to compute the
anomaly:
\begin{equation}
  \label{eq:38}
  \anom= \frac {e}{\pi}\pxp A_-(x) \F^2(r)
\end{equation}
This result is a product of  two factors: $\anom_{1+1}$ -- coming from
$(t,z)$ part of the computation and the contributions from transversal
directions -- $\F(x_\perp)$. For $\anom_{1+1}$ we have:
\begin{equation}
  \label{eq:39}
  \anom_{1+1} = \frac {e}{\pi}\pxp A_-(x)
\end{equation}
This coincides with the usual result for two-dimensional anomaly
(compare~\cite{Jackiw, Naculich}). The standard result is usually expressed in
a more symmetric way: 
\begin{equation}
  \label{eq:2}
  \tilde \anom_{1+1} = \frac {e}{2\pi} (\pxp A_- - \pxm A_+) 
  = - \frac{e}{8\pi}\epsilon^{ab}F_{ ab }
\end{equation}
It differs from eq.~(\ref{eq:39}), which is just the well-known 
fact that one is
free to add any local counterterm ($\int d^2x A^2(x)$ in this case) to the
action and thus modify the anomaly from the form~(\ref{eq:39}) to the
form~(\ref{eq:2}).

As a consistency check, the total charge non-conservation (as viewed from
infinity) should coincide with that of 1+1 dimensional anomaly.  This will be
the integral of~(\ref{eq:38}) over the transversal space:
\begin{equation}
  \label{eq:41}
  \int d^2 x_\perp \anom =  \anom_{1+1} = - \frac e {8\pi} \epsilon^{ab} F_{ab}
\end{equation}
(we will not distinguish between the form~(\ref{eq:2}) and~(\ref{eq:39})).
Thus, variation of $S_{eff}$ in the bulk under the gauge transformation $A_\mu
\rightarrow A_\mu + \frac{1}{e} \partial_\mu \Lambda$ will be:
\begin{equation}
  \label{eq:42}
  \delta_\Lambda S_{eff} = - \int d^4x\, \Bigl (\partial_\mu j^\mu_\zm \Bigr )\Lambda
  = \int d^4x\,\anom\, \Lambda  
\end{equation}
Substituting the result~(\ref{eq:38}) into~(\ref{eq:42}) and using polar
coordinates in the transversal direction we get:
\begin{equation}
  \label{eq:43}
\delta_\Lambda S_{eff}= -\frac e {8\pi} \int
  dr\, d\varphi\, \underbrace{r\F^2(r)} \int  d^2 x^a \, \epsilon^{ab}
  F_{ab} \Lambda (x) 
\end{equation}
To extract the value of the $\rho(r)$ function from eq.~(\ref{eq:43}), note
that it follows from eq.~(\ref{eq:S_eff3}) that
\begin{equation}
  \label{eq:44}
  \delta_\Lambda S_{eff} = - \frac {e} {8\pi}\int\limits_{M_4} 
  dr d\varphi d^2 x^{int}  \rho^\pr(r) \epsilon^{ab} F_{ab} \Lambda (x)
\end{equation}
where we've used the expression~(\ref{eq:dual5}) for the delta-function 
embedding the string's world-sheet into the bulk space. From~(\ref{eq:44}), we
see that the underbraced expression in~(\ref{eq:43}) plays the role of $\frac
1 {2\pi} \rho^\pr(r)$. Indeed, $r\F^2(r) = 0$ as $r\rightarrow 0$ and to
$\infty$. The normalization condition~(\ref{eq:12}) is enough to satisfy the
property~(\ref{eq:rho4}). So, the natural claim would be that:
\begin{equation} 
  \label{eq:45}
 \rho(r) = \dfrac 1 {2\pi} \int\limits_0^r d\sigma \sigma \F^2(\sigma)
\end{equation} 

We see, that it is possible to write an expression for the effective action in
the whole bulk, taking into account the  structure of the zero modes of 
the theory.
Naively, one might have expected the axion string
profile  $f(r)$ to be  a candidate for the
$\rho$-function. It is the function $\F(r)$ however, which describes the 
zero mode
behavior in the bulk, so eq.~(\ref{eq:45}) tells us that the bump-form is
related to the profile of the zero modes $\F(r)$, rather than to the profile
of the scalar field. 

\section{Discussion}
\label{sec:discussion}

As would be expected physically, the analysis here shows that the
anomalous divergence of the current on the axion string is spread
out over a region whose size is determined by the radial wave function
of the fermion zero modes. This requires a corresponding dependence
on the  zero mode profile in the non-zero mode contribution to
the action which is reflected in the bump function and the
modification to the action given in~(\ref{eq:S_eff3}).

In the mechanism for M-theory fivebrane anomaly cancellation 
of \cite{FHMM} a bump function also appears in the modification
of the Bianchi identity for the four-form field strength,
\begin{equation}
  \label{eq:69}
  d G_4/{2\pi} = d\rho \land e_4/2
\end{equation}
where $e_4/2$ is the global angular form. A naive extension of
the analysis of  this paper suggests that this bump
function should be expressed in terms of the zero mode
radial wave function as
\begin{equation}
  \label{eq:68}
  d\rho(r) = r^4 \F^4(r) dr
\end{equation}
The radial profile $\F(r)$ has power law fall-off away from
the fivebrane as compared to the exponential fall-off for
the axion string. For example, the gravitino variation has
the form 
\begin{equation}
  \label{eq:65}
  \begin{array}{rcl}
    \delta \Psi^\mu & \propto &\left ( 1 + \frac {1}{r^3} \right
    )^{-3/2}\cdot   \frac {1} {r^3}\\
    \delta \Psi^m & \propto & \left ( 1 + \frac {1} {r^3} \right )^{-7/6}\cdot 
    \frac {1}{r^3}\\
  \end{array}
\end{equation}

In addition, cancellation of the M-theory fivebrane anomaly 
requires non-local  modification of the Chern-Simons couplings
$\int C_3 \wedge G_4 \wedge G_4$. Our analysis
suggests that these modification should not be viewed as corrections
to the fundamental action of M-theory, but rather arise from the
splitting of the action into a zero-mode action on the fivebrane and
an effective bulk action away from the fivebrane. A microscopic
derivation of these modifications would be desirable, but the
axion string we have discussed is not a good model for understanding
these modifications. The axion string in four
dimensions does have an uncancelled normal bundle anomaly of
$p_1(N)/6$, but this is trivially canceled by a local counterterm
in the world-sheet effective action \cite{Witten,Becker}.

\end{document}